\documentclass[preprint, a4paper,5+,10pt,twoside]{elsarticle}
\usepackage[margin=1in]{geometry}
\usepackage{graphicx}
\usepackage{epstopdf}
\usepackage{lineno,hyperref}
\usepackage{multicol}
\usepackage{textcomp}

\begin{document}
     
\begin{frontmatter}

\title{tACS Facilitates Flickering Driving by Boosting Steady-State Visual Evoked Potentials}


\author[tsinghua]{Bingchuan Liu\fnref{fn1}}
\ead{louislbc@gmail.com}
\author[tsinghua]{Xinyi Yan}
\ead{yanxy14@mails.tsinghua.edu.cn}
\author[cams]{Xiaogang Chen}
\ead{chenxg@bme.cams.cn}
\author[cas]{Yijun Wang\corref{cor3}}
\ead{wangyj@semi.ac.cn}
\author[tsinghua]{Xiaorong Gao\corref{cor1}}
\ead{gxr-dea@mail.tsinghua.edu.cn}
\address[tsinghua]{Department of Biomedical Engineering, School of Medicine, Tsinghua University, Beijing, 100084, China}
\address[cams]{The Institute of Biomedical Engineering, Chinese Academy of Medical Sciences and Peking Union Medical College, Tianjin, 300192, China}
\address[cas]{State Key Laboratory on Integrated Optoelectronics, Institute of Semiconductors, Chinese Academy of Sciences, Beijing 100083, China}
\cortext[cor1]{To whom correspondence may be addressed}

\begin{abstract}
     There has become of increasing interest in transcranial alternating current stimulation (tACS) since its inception nearly a decade ago. tACS in modulating brain state is an active area of research and has been demonstrated effective in various neuropsychological and clinical domains. In the visual domain, much effort has been dedicated to brain rhythms and rhythmic stimulation, i.e., tACS. However, little is known about the interplay between the rhythmic stimulation and visual stimulation. Here, we used steady-state visual evoked potential (SSVEP), induced by flickering driving as a widely used technique for frequency-tagging, to investigate the aftereffect of tACS in healthy human subjects. Seven blocks of 64-channel electroencephalogram were recorded before and after the administration of 20-min 10-Hz tACS, while subjects performed several blocks of SSVEP tasks. We characterized the physiological properties of tACS aftereffect by comparing and validating the temporal, spatial, spatiotemporal and signal-to-noise ratio (SNR) patterns between and within blocks in real tACS and sham tACS. Our result revealed that tACS boosted the 10-Hz SSVEP significantly. Besides, the aftereffect on SSVEP was mitigated with time and lasted up to 5 min. Our results demonstrate the feasibility of facilitating the flickering driving by external rhythmic stimulation and open a new possibility to alter the brain state in a direction by noninvasive transcranial brain stimulation.
  
\end{abstract}
\begin{keyword}
\texttt   Noninvasive transcranial brain stimulation \sep Transcranial alternating current stimulation (tACS) \sep Steady-State visual evoked potentials (SSVEP) \sep Cortical excitability \sep  Aftereffect
\end{keyword}
\end{frontmatter}
\section{Introduction}
Over the last few decades, noninvasive transcranial brain stimulation gains increasing attention in the field of human neuroscience, because its emergence provides the possibility to modulate cortical excitability of brain by external methods. Among all noninvasive stimulation methods, transcranial alternating current stimulation (tACS), is a novel neuromodulation technique by applying sinusoidal stimulations on the scalp and thereby contributes to the alteration of brain oscillation\cite{yavari2017basic}. Previous studies have demonstrated that tACS has the capacity to entrain endogenous oscillations in a frequency-dependent manner\cite{kanai2008frequency-dependent, zaehle2010transcranial, merlet2013from, neuling2013orchestrating, helfrich2014entrainment}. For example, in the alpha band, Zaehle et al. observed that the real tACS increased the endogenous alpha oscillation power in the occipital cortex for the first time\cite{ zaehle2010transcranial}. Merlet et al. obtained similar results by building an integrated model to simulate the effects of tACS on scalp EEG\cite{merlet2013from}. Besides electrophysiological methods, by using functional imaging such as fMRI a few studies found that tACS influenced BOLD signals as a function of stimulation frequency, intensity and task condition\cite{vosskuhl2016bold, cabral2016transcranial}. Some studies found that the duration of the aftereffect of tACS ranging from 1 min to 70 min varied with the duration of stimulation\cite{bergmann2009acute, struber2014antiphasic, helfrich2014selective, kasten2016sustained}. Furthermore, it seems that the effect of tACS is also influenced by the brain state\cite{neuling2013orchestrating, ruhnau2016eyes}. 

Meanwhile, plenty of behavioral studies also found that appropriate tACS protocol would influence the performance of tasks, including motor tasks such as voluntary movements\cite{antal2008comparatively, pogosyan2009boosting, joundi2012driving}, memory functions \cite{reinhart2019working, marshall2006boosting, kirov2009slow}, functional connectivity\cite{cabral2016transcranial} and higher-order cognition such as fluid intelligence \cite{santarnecchi2013frequency}. The ability to modulate ongoing oscillations makes tACS become an ideal tool to study the causal link between brain oscillations and cognitive functions. Consequently, tACS is a promising new method for the treatment of oscillation-related disorders, such as Alzheimer’s disease, epilepsy, Parkinson’s disease, and schizophrenia.\cite{uhlhaas2012neuronal, brittain2013tremor, ahn2019targeting}. 

The effect of tACS on visual system is less reported. Kanai and co-workers firstly finished a series of researches on the impact of occipital tACS on phosphene and demonstrated that the effect was influenced by the external environment, like illuminated condition\cite{kanai2008frequency-dependent, laczo2012transcranial}. Further studies doubted the direct effects of tACS on visual cortex because tACS could also induce retinal cutaneous activation\cite{schutter2016cutaneous}. However, more behavioral researches have reported the modulation of occipital tACS on visual cortex\cite{laczo2012transcranial, brignani2013transcranial,herring2019low-frequency}. For example, Lacz\'o et al. found that gamma tACS significantly influenced contrast discrimination that was based on V1 activities\cite{laczo2012transcranial}. However, apart from behavioral and phosphene studies, there are few existing studies in the visual domain, specifically on visual evoked potentials (VEPs). Related work is that Philipp et al. investigated the concurrent effects of 7 and 11Hz tACS on matched steady-state responses (SSR) by removing the artifacts in MEG recordings and suggested frequency-specific effects of tACS\cite{ruhnau2016flicker-driven}.

Evoked potentials in the visual system, especially steady-state visual evoked potentials (SSVEPs), are widely utilized in the scientific study of visual attention \cite{morgan1996selective, muller1998the, muller2003sustained} and applied settings of clinical diagnose \cite{schomer2012niedermeyer, medeiros2016ngoggle}. SSVEPs are periodic signals that are neural responses to flickering visual stimulation at specific frequencies, generally ranging from 3 Hz to 60 Hz\cite{regan1989human, herrmann2001human}. Some properties of SSVEP, such as high signal-to-noise ratio (SNR)\cite{vialatte2010steady-state}, make SSVEP a prime candidate for engineering applications, for instance, brain-computer interface (BCI)\cite{gao2003a, chen2015high-speed}. 

In the present study, we focus on the interaction between tACS and periodic visual stimulation, specifically the effect of tACS on SSVEP. First, we suppose that weak tACS is able to penetrate the skull and stimulate the visual cortex, as is evidenced by previous study\cite{ozen2010transcranial}. This assumption constitutes the basis for the common findings that tACS can modulate visual cortex\cite{kanai2008frequency-dependent, ruhnau2016flicker-driven, laczo2012transcranial, brignani2013transcranial}. Second, SSVEP is a special type of oscillatory brain response, and its derivation might be related to the brain rhythms \cite{notbohm2016modification, capilla2011steady-state}. Since tACS has the capacity to modulate brain rhythms \cite{ zaehle2010transcranial, bergmann2009acute, kasten2016sustained} and the frequency-specific property characterizes both tACS and SSVEP, we assume that tACS might influence SSVEP. Because of the fact that the power of SSVEP is highest in the alpha band (8--12Hz), we chose the central frequency of the band, 10 Hz, as the stimulus frequency of SSVEP. As recommended by \cite{herrmann2013transcranial} to choose the match frequency, 10 Hz is chosen both for tACS and SSVEP in this study. This is in line with the previous studies\cite{merlet2013from, zaehle2010transcranial} that matched frequency of tACS with brain oscillation might lead to resonance and the modulated effect is most obvious. In sum, we hypothesize that appropriate 10-Hz tACS might influence 10-Hz SSVEP.

\section{Materials and Methods}
\subsection{Participants}
    Twelve healthy subjects (age: 20.5$\pm$2.2 years, six males and six females) participated in this study. They were all paid volunteers randomly recruited from Tsinghua University. All the participants met the following requirements (1) right-handedness (2) normal or corrected-to-normal vision (3) no attention-deficit or hyperactivity disorder (4) no history of epileptic seizures or other neuropsychiatric disorders (5) no intake of caffeine, alcohol or medication (6) no fatigue prior to the experiment (7) no history of brain injury or intracranial implantation. Full written informed consent was given by the subjects at the beginning of the experiment. This study was approved by institutional review board of Tsinghua University (No. 20190021) and was under the declaration of Helsinki.
\subsection{Stimuli}
    In this study, we adopted a flickering square to evoke SSVEP and the visual flicker was generated by PsychoToolbox \cite{brainard1997psychophysics} at a frequency of 10 Hz. The size of the flicker was 250$\times$250 pixels (visual angle: 6.5$^{\circ}$) and lay at the center of a 27-inch LCD screen (LG 27GK750F, refresh rate: 60Hz, resolution: 1920$\times$1080 pixels). Initially, a photocell test was conducted to measure photoelectric responses elicited by the flicker. Besides, a frame stability test was performed to record the frame interval of each trial. Both tests were passed to ensure the preciseness of visual presentation before carrying out the experiment.
\subsection{Procedure}
    We chose a within-subject design to exclude confounding variables for this study. All subjects took part in two experiments, i.e., real tACS and sham tACS in counterbalanced order on different days. For each subject, the start time of experiments was fixed at either 9 a.m. or 7 p.m. to ensure the participants to have a clear and comparable mental state for both experiments.
    \par For each experiment, subjects underwent a practice session, a baseline session, a tACS session, and a post-tACS session. In both the practice session and baseline session, a block of SSVEP task was performed. The practice session was set for participants to be familiar with the experiment and data collected from this session were not used for analysis. The baseline session provided us the data as a baseline prior to tACS stimulation. In the tACS session, participants received real tACS or sham tACS depending on the type of experiment and no blocks of the task were performed. The post-tACS session came instantly after the tACS session and consisted of 5 blocks of tasks. 
    \par The block of SSVEP task consisted of 80 trials, each of which comprised 2-s flickering and 1-s black screen interval for rest. To avoid artifacts and alpha oscillation from eye blink \cite{toscani2010alpha}, subjects were required to stare at the center of the flicker without blinking. Meanwhile, 12 Go/No-Go tasks were randomly inserted in the SSVEP tasks to measure behavioral performance. Each task block lasted approximate 4 min 30 s in total, including the SSVEP task and Go/No-Go task. During the task block, participants were advised to sit calmly and not to move. Participants have a 2-min break between two adjacent blocks. During the 2-min break, participants were instructed to rest calmly with eyes open. The whole procedure was carefully timed.
\subsection{Transcranial Alternating Current Stimulation}
     To achieve spatially high-definition recording and stimulation, we combined a tACS device (StarStim, Neuroelectrics, Inc., Barcelona, Spain) with the 64-channel Electro-Cap. Transcranial alternating current was therefore delivered to two PiStim stimulating electrodes ($\pi$ cm$^{2}$  area, Neuroelectrics, Inc.), which lay respectively in the sagittal vicinity of Oz and Cz recording electrodes underneath the Electro-Cap. Previous work of finite-element simulation demonstrated convincing evidence that this montage ensured a maximum of current densities in occipital cortex\cite{neuling2012finite-element}. The stimulating current was set constant at an intensity of 0.65 mA according to the previous study\cite{neuling2015friends}. The frequency of tACS was at 10 Hz in a sinusoidal manner. No DC offset was added to the AC waveform.
    \par To alleviate electrochemical sensations, the current in real tACS ramped up for 10 s at the beginning of the stimulation, maintained stable and finally ramped down for 10 s at the end of stimulation. For the purpose of blindness, the sham tACS also brought about comparable sensations by ramping up the current for 10 s and immediately ramping down for another 10 s. The duration of stimulation was 20 min \cite{neuling2013orchestrating} for both the real and sham tACS. The impedance of the stimulation electrode was kept below 2k$\Omega$ before the start of the practice session and it was double-checked at the beginning of tACS session. During the tACS session, participants are asked to rest calmly with eyes open.
\subsection{Go/No-Go Task}
    We designed a Go/No-Go task in each block of SSVEP task to keep the subject alert and also measured the subject's capacity for response control and sustained attention\cite{nosek2001go}. In the Go/No-Go task, subjects were required to press the SPACE button (Go response) as quickly as possible when presented with target stimuli and not to respond (No-Go response) when presented with distractor stimuli. The target stimuli and distractor stimuli are defined as follows. As mentioned before, there were 7 blocks of SSVEP tasks in total (2 pre-tACS blocks and 5 post-tACS blocks) and the Go/No-Go task was randomly inserted in each block. During the Go/No-Go task, one of the numbers from 1 to 12 was presented on the screen in random order. The target stimulus was the prompt of odd number in the odd blocks (1st, 3rd, 5th, 7th) whereas it was the even number in the even blocks (2nd, 4th, 6th). The prompt of number was presented for 0.5 s and responses that were later than 1.5 s after the prompt appeared were regarded as failure responses. 
\subsection{Visual Analog Scale}
    At the end of each session, participants were evaluated by a Visual Analog Scale (VAS)\cite{knotkova2016textbook} to report their subjective sensation. For the practice, baseline and post-tACS sessions, a VAS of fatigue evaluation was performed. Ten scores ranging from 1 to 10 indicated the extent of fatigue. Specifically, 10 denoted no fatigue and highly focused attentiveness. Scores from 9 to 1 were split into three intervals of slight fatigue, moderate fatigue, and severe fatigue, respectively. For the tACS session, a VAS of discomfort evaluation was performed instead. At the end of stimulation, subjects rated their sensation of discomfort to scores from 2 to 10. 1 denoted no discomfort and scores from 2 to 9 were split into three intervals of slight discomfort, moderate discomfort, and severe discomfort, respectively. 
\subsection{Data Acquisition}
    Three types of EEG data were recorded, including SSVEP data, 2-min rest data between two consecutive blocks and 20-min EEG data under stimulation. All these data were recorded by SynAmps2 (Neuroscan Inc., Charlotte, USA) at a sampling rate of 1000Hz. The 64-channel recording montage was aligned according to the international 10-10 system and the vertex Cz electrode was used as the reference. The impedance of all recording electrodes were kept below 10k$\Omega$ before the experiment and were double-checked periodically throughout the experiment. The experiment was performed in a sound-attenuated and electrically shielded room (CA acoustics, Beijing, China). All lights were turned off in the room during the experiment.
    \par Five EEG data (a break after the practice session and four breaks in the post-tACS session) were also collected during the 2-min rest. Apart from the EEG data, two behavioral data, i.e., the VAS data and Go/No-Go data including response type and reaction time were collected for behavioral analysis.
\subsection{Data Analysis}
\subsubsection{EEG data}
    We firstly preprocessed the raw EEG data by visual inspection in EEGLAB (Salk Institute, La Jolla, CA, USA). SSVEP trials contaminated with movement artifacts were rejected, leaving 98.98$\pm$0.28 \% of the original data for analysis. EEGs were then band-passed filtered from 3.5 to 100 Hz with an infinite impulse response (IIR) filter using the standard eegfilt() function in EEGLAB. For each block, 2-s SSVEP epochs were then extracted and averaged to one trial to enhance the SNR of SSVEP. The averaging procedure in the time domain before frequency-domain analysis was in accord with previous study\cite{wieser2014fearful}.
    \par Following preprocessing, we then determined the SSVEP power for each averaged trial in the frequency domain. The SSVEP power was obtained by calculating the square of the Fourier coefficient of the 2-s trial at the stimulus frequency, i.e., 10 Hz in the study. This procedure was performed to all channels in one block, and all blocks in either real tACS or sham tACS. Considering in the SSVEP powers there existed large inter-subject variation\cite{silberstein1990steady-state}, we applied a procedure of SSVEP power normalization according to the previous study\cite{silberstein2000steady-state}. For each subject, an SSVEP power normalization factor was calculated by averaging the SSVEP powers across all channels and all blocks. Since the impedance of the trials in real tACS and sham tACS could be different, we obtained the respective normalization factor in real tACS and sham tACS. The normalized SSVEP power was formed by dividing the SSVEP power in real tACS and sham tACS by its associated SSVEP power mean factor. 
     \par To measure the aftereffects induced by tACS session, the change in SSVEP power was obtained by subtracting the SSVEP power in the baseline session from that of the post-tACS session. This procedure was applied to each subject and each channel in both real tACS and sham tACS. A planned comparison was carried out by performing Mann-Whitney U-test to the change in SSVEP power in real tACS versus sham tACS. In this manner, we could obtain a p-value at each electrode and post-tACS time level. The change in SSVEP power and its corresponding statistical significance were topographically mapped using cubic spline interpolation in EEGLAB. To better illustrate the significance of the change, p values greater than 0.05 in the topography were colored white. 
    \par To characterize the change in SSVEP power, we further investigated the SSVEP envelope of both conditions in the baseline and post-1 sessions. Since Oz electrode lay in the central occipital region where SSVEP signals were predominantly distributed, it was taken as the representative electrode for the analysis of SSVEP. SSVEP time series were initially normalized by their respective SSVEP power normalization factor for each subject and each tACS condition. Then narrow-band filtering (center frequency:10 Hz, bandwidth: 0.2 Hz) with an 8-order Butterworth filter was performed to each trial. In the implementation, zero-phase forward and reverse filtering was achieved using the filtfilt() function in EEGLAB. The SSVEP envelop was formed by applying a Hilbert transform to the filtered trials and then taking the modulus of complex numbers \cite{zhang2018a}. For each subject and each tACS condition, the SSVEP envelope was obtained in this fashion. For real tACS and sham tACS, the mean and standard error of SSVEP envelop were calculated respectively and plotted. A paired t-test was applied to the change in the envelope (post minus pre) between real tACS and sham tACS at each time point. The topographic maps of change in SSVEP power (the change in real tACS minus the change in sham tACS) at different stages within 2000 ms (0$\sim$1275 ms, 1275$\sim$1700 ms, 1700$\sim$2000 ms) were generated for comparison.
    \par In parallel to the envelope, we investigated the dynamic phase properties of SSVEP in both conditions by calculating the phase lock value (PLV) \cite{lachaux1999measuring,aydore2013a} as follows.
    \vspace{5pt}
    
    {\centering \textit{PLV$_{t}$} = $\frac{1}{N}$$\mid$$\sum_{n=1}^N$ exp(\textit{j}$\theta$(t,n))$\mid$
    
    }
    
    \vspace{5pt}
    where N denoted the number of trials and $\theta$(t,n) denoted the phase difference $\phi$$_{1}$(t,n) -- $\phi$$_{2}$(t,n). Since SSVEP was phase-locked to the stimulus\cite{vialatte2010steady-state} and the stimulus was generated using the sampled sinusoidal stimulation method \cite{manyakov2013sampled, Chen2014A}, we used a sinusoidal signal (frequency: 10Hz, initial phase: 0$^{\circ}$) as a reference signal. The $\phi$$_{1}$(t,n) and $\phi$$_{2}$(t,n) were estimated by applying a Hilbert transform to the trial after the trial was band-pass filtered as mentioned above. This procedure was applied to each block per subject for both conditions of real tACS and sham tACS.
    \par In addition, since SSVEP contained multiple harmonics in the frequency domain\cite{norcia2015the}, we further studied the time-frequency characteristics of SSVEP on the Oz electrode for the baseline and post-1 sessions. The time-frequency representation was calculated from single-trial SSVEP epoch using the short-time Fourier transform (window length: 512, overlap: 500) and then it was averaged for each subject under each condition. The change in time-frequency representation was obtained by subtracting the values of the baseline session from the post-1 session for real tACS and sham tACS, respectively. And finally the difference between the change in real tACS and sham tACS was calculated, and the spectrogram was plotted to illustrate the change in SSVEP harmonics. The Mann-Whitney U-test was performed to the change in time-frequency representation between real tACS and sham tACS at each pixel of the spectrogram. The yielded statistical significance was plotted and p-values greater than 0.05 were colored white for illustrative purpose.
     \par To delve into the aftereffect of tACS on SSVEP, we analyzed the ongoing change in single-trial SSVEP power within a block. We chose the remaining trials, i.e. the first 78 trials after artifact rejection (on average two trials were rejected), to calculate their respective single-trial SSVEP power for the baseline and post-1 sessions in both tACS and sham conditions. For comparison, SSVEP power ratio was defined as follows to characterize the change relative to its associated baseline. Values in the post-1 session were divided by the mean value of baseline session for real tACS and sham tACS, respectively, and the values in baseline session were then divided by their mean. The SSVEP power values before the division procedure were used for statistical analysis. The sequence of single-trial SSVEP power was then averaged across subjects and fitted with linear regression to quantify the dynamics of the aftereffect. To unveil the changes in SSVEP power at a global scale, all the 78 trials were divided into 3 stages, i.e., early stage (1$\sim$26), middle stage (27$\sim$52) and late stage of a block (53$\sim$78). The spatial patterns of SSVEP power were plotted for the post-1 session in the real tACS condition.
     \par SSVEP is based on frequency coding and its available signal component can be evaluated by the metric of signal-to-noise ratio (SNR). To characterize wide-band noise and the contribution of harmonics to the signals, the SNR (in decibels, dB) is defined as follows\cite{zhang2018a}
     \begin{equation}
    SNR=10log_{10}\frac{\sum^{k=Nh}_{k=1}P(k\cdot f)}{\sum^{f=f_s/2}_{f=0}P(f)-\sum^{k=Nh}_{k=1}P(k\cdot f)}
    \end{equation}
    where $P(f_n)$ is the power spectrum at frequency $f$, $Nh$ is the number of harmonics, and $f_s$ is the sampling rate. From the perspective of brain-computer interface (BCI), we calculated the mean SNR with five harmonics ($Nh=5$) using data from nine parietal and occipital electrodes (Pz, PO3, PO4, PO5, PO6, POz, O1, Oz and O2). This procedure was applied to single trials in the three stages (initial:1$\sim$26 trials; middle: 27$\sim$52 trials; late: 53$\sim$78 trials), yielding the mean SNR values for each stage of the baseline and post-1 session in both real and sham tACS conditions. The Shapiro-Wilk test of normality was evaluated to the studentized residuals to assess normality of the SNR data. A planned paired t-test was then applied to obtain the statistical significance of the comparison between baseline session and post-1 session for real tACS and sham tACS, respectively.
    \par Furthermore, we analyzed the 2-min EEG data recorded during the break between two consecutive blocks. EEG epochs with movement artifacts and eye blinks were rejected by visual inspection. After preprocessing, two metrics, i.e., the relative alpha power and relative 10-Hz power were then calculated. Specifically, the relative alpha power was obtained by computing the ratio of alpha band (8$\sim$12 Hz) power spectral to the whole-band power spectral and relative 10-Hz power was obtained in a similar fashion. To overcome the non-gaussian distribution of the data, we took a logarithm of the resultant value and multiplied it by 20 (in decibels). In accord with the SNR analysis, the nine parietal and occipital electrodes were selected for analysis and the associated metric values were then averaged by channel. This procedure was conducted for the 5 rest data (1 pre-tACS and 4 post-tACS) in both real and sham tACS. The procedure of two-way repeated measures ANOVA and the post-hoc t-test were applied for statistical analysis on the two metrics.
\subsubsection{Behavioral data}
    For the Go/No-Go task, we calculated the sensitivity index d' in each block for both real tACS and sham tACS. According to the signal detection theory, the d' was given as follows\cite{stanislaw1999calculation}.
    \vspace{5pt}
    
    {\centering d' = Z(hit rate) -- Z(false alarm rate)
    
    }
    
    \vspace{5pt}
    Where Z(p) denoted the z score of the probability. Subjects who achieved 100\% false alarm rate (i.e. press button for all distractor stimuli) one block were considered as a failure to remember the target stimuli and were removed for analysis. In addition, we further extracted the reaction time (RT) when the subject correctly hit the target. Repeated measures of ANOVA was performed to both d' and RT for real tACS versus sham tACS. For the data of VAS, the scores of discomfort and fatigue were extracted and applied with repeated measures ANOVA respectively for both conditions. All the previous procedures were processed in MATLAB 2018b (The MathWorks Inc, Natick, MA, USA) and SPSS Statistics 20 (IBM, Armonk, NY, USA). All the data were expressed as mean$\pm$ s.e. (standard error of the mean).
%
\section{Results}

 \subsection{Changes in spatial and temporal profile of SSVEP} 
\begin{figure}[h!]
 \begin{center}
   \includegraphics[width=10cm]{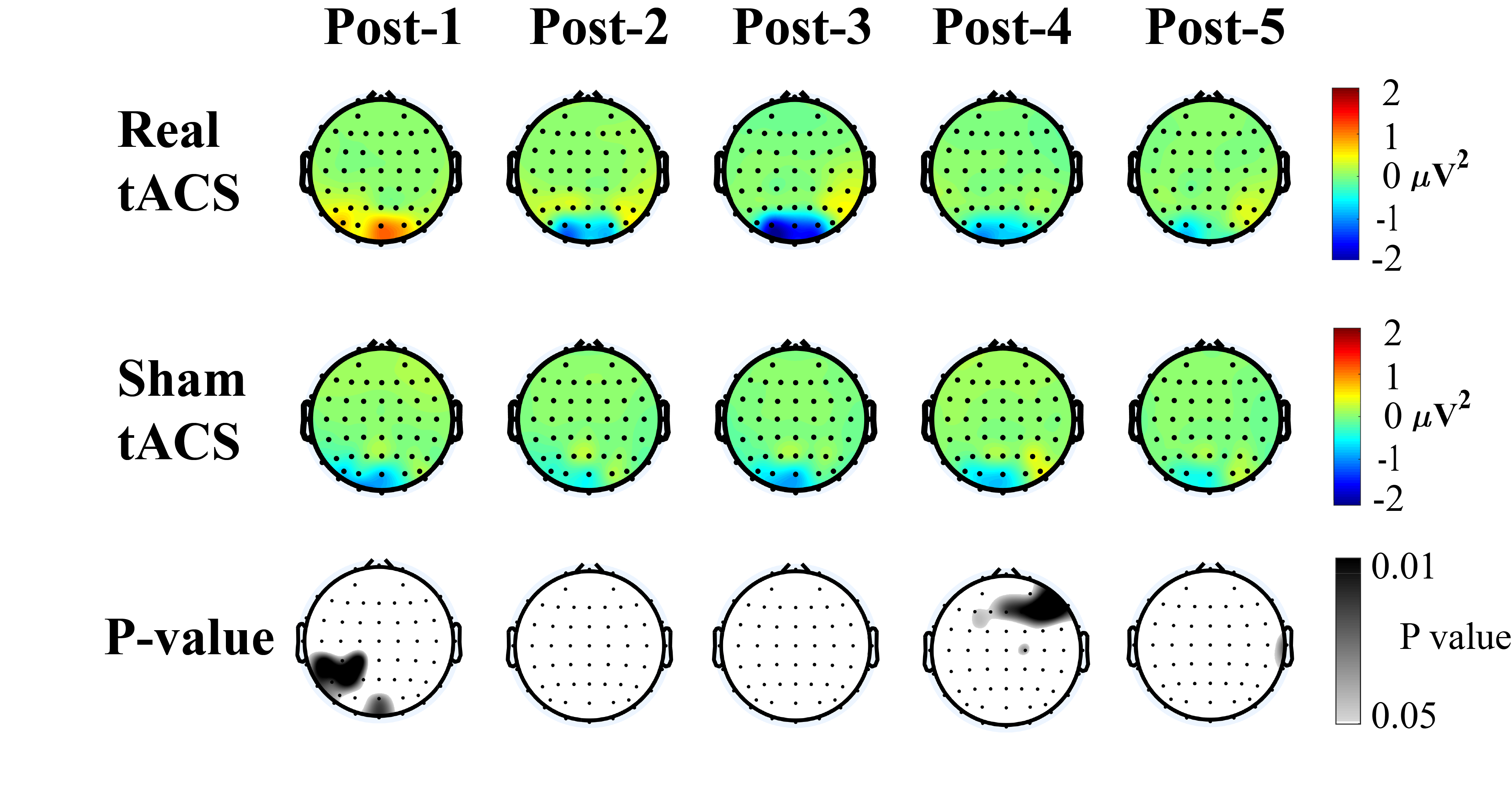}
   \caption{Topographic maps of change in SSVEP powers and their corresponding statistical significance. The first two rows depict the grand-averaged change in SSVEP power relative to the baseline session during five post-tACS blocks for real tACS and sham tACS, respectively. The warmer color indicates an increase in SSVEP power and cooler color indicates a decrease in SSVEP power. The third row depicts the statistical significance between the two conditions from Mann-Whitney U-test. P values greater than 0.05 were colored white. Note that the central occipital region showed a significant increase in SSVEP power in the post-1 block.}    
   \label{fig:1}
 \end{center}
\end{figure}

\begin{figure}[h!]
 \begin{center}
   \includegraphics[width=7cm]{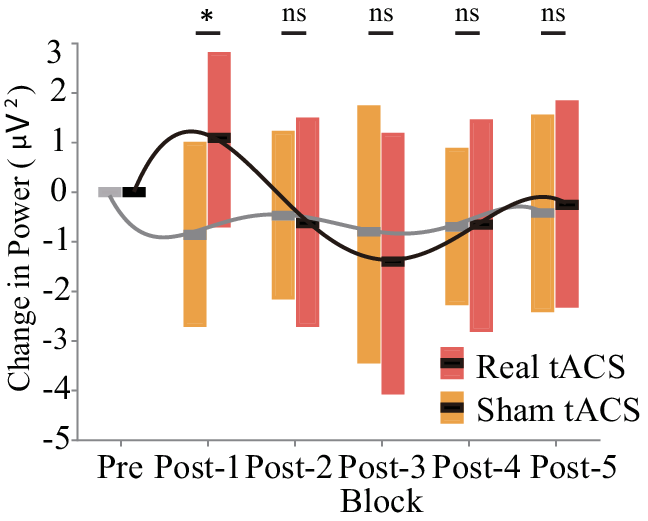}
   \caption{Changes in SSVEP power on Oz electrode relative to the baseline session (Pre) following the administration of tACS. Only the post-1 session reveals statistical significance (p\textless0.05).}    
   \label{fig:2}  
 \end{center}
\end{figure}

     For each block and condition, the global changes in SSVEP power relative to the baseline session were illustrated in Fig.\ref{fig:1}. In the post-1 block of the real tACS condition, we could noticeably observe a prominent increase in SSVEP power on the occipital and posterior parietal region. In contrast, a focal decrease was found in the occipital region of sham tACS condition in the post-1 block. By comparing these two conditions, the regions of central occipital and posterior temporal lobes both revealed statistical significance ($p<0.05$), as indicated by the topographic map of statistical significance in Fig.\ref{fig:1}. During the blocks from post-2 to post-5, the occipital SSVEP powers tended to decrease relative to the baseline session in both real and sham tACS conditions. No statistical significance was found following the post-1 block, except for the right frontal region in the post-4 block. Specifically, a detailed version of Oz electrode was illustrated in Fig.\ref{fig:2}. In accordance with the result in Fig.\ref{fig:1}, the relative SSVEP power of Oz electrode showed statistically significant only in the post-1 session, $p = 0.014$. 
     \par
\begin{figure}[h!]
 \begin{center}
   \includegraphics[width=12cm]{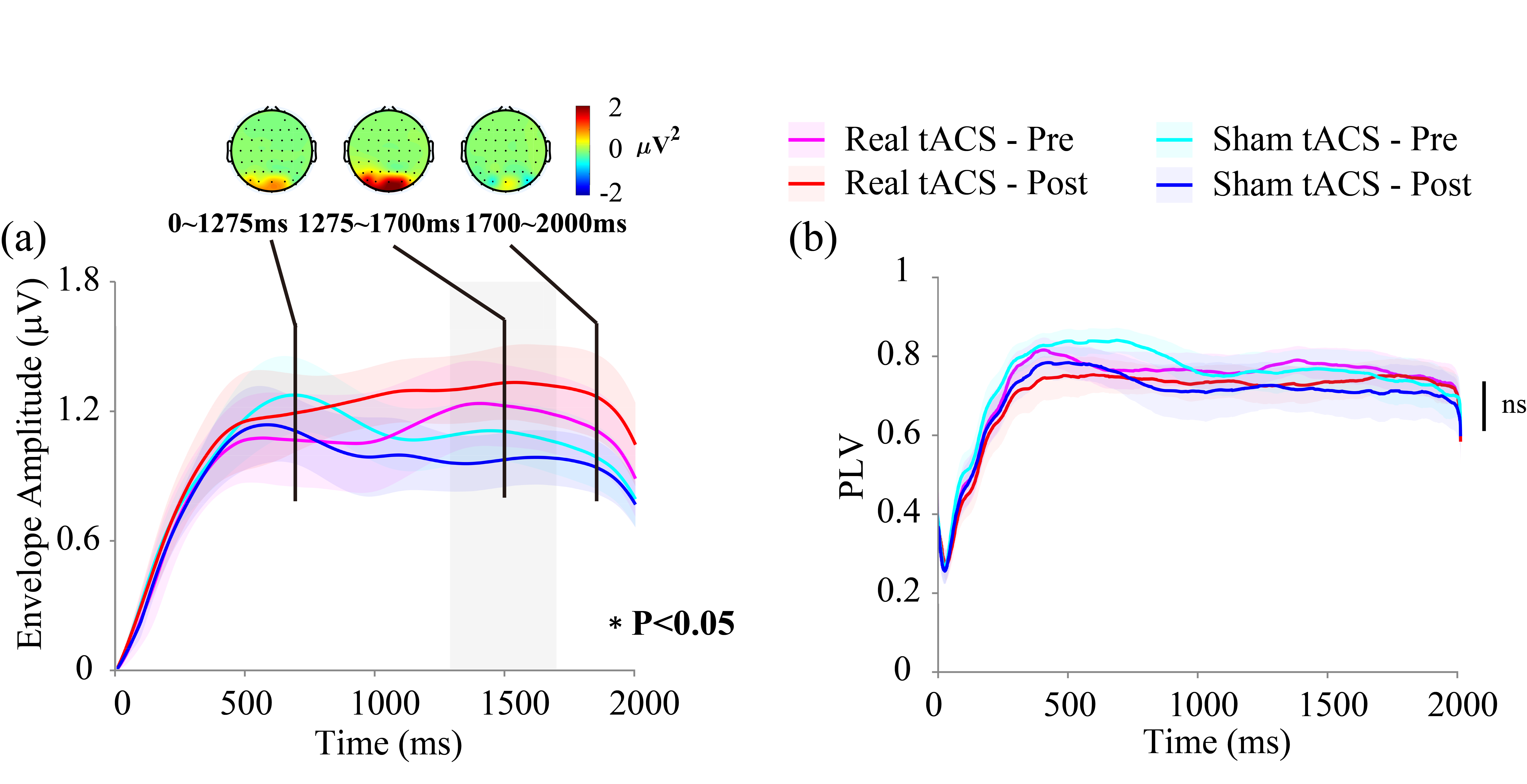}
   \caption{Temporal characteristics of grand-averaged SSVEP epochs for real tACS and sham tACS. The baseline session (Pre) and post-1 session (Post) were analyzed for comparison. The left panel (a) depicts envelope of narrow-band SSVEP signals and the right panel (b) depicts the phase lock value (PLV) of SSVEP relative to sinusoidal signal. The left upper panel illustrates spatial pattern of the changes in SSVEP power (the change in real tACS minus the change in sham tACS) during three stages (0$\sim$1275 ms, 1275$\sim$1700 ms, 1700$\sim$2000 ms). Gray shaded areas indicate statistical significance (p\textless0.05, 1275$\sim$1700 ms) between the change in two conditions (change in real tACS versus change in sham tACS) and colored shaded areas indicate standard error.}    
   \label{fig:3}  
 \end{center}
\end{figure}

 The envelope and phase dynamics of SSVEP on Oz electrode were illustrated in Fig.\ref{fig:3} (mean$\pm$ se). As indicated by the envelope result (Fig.\ref{fig:3}a), the overall amplitude of envelope in the real tACS condition was elevated following tACS. This was in stark contrast with the sham tACS condition, where the amplitude of envelope exhibited a downward trend. Notice that at the start of SSVEP progression, there was no marked distinction between the real tACS and sham tACS group. However, in the later stage of the steady-state stage ($1275\sim1700$ ms), the distinction of change in SSVEP envelope (change in real tACS versus change in sham tACS) became significant ($p<0.05$), as indicated by the grey shaded area. This was also in line with the spatial pattern of SSVEP power (Fig.\ref{fig:3} top). Notably, we could observe a prominent increase in SSVEP power in occipital region during 1275 to 1700 ms, compared to its preceding stage ($0\sim1275$ ms) and subsequent stage ($1700\sim2000$ ms). On the other hand, the phase result (Fig.\ref{fig:3}b) revealed a marked increase in PLVs when SSVEP achieved steady state, with a value of $0.744\pm0.032$ for $500\sim2000$ ms. However, no significant difference was found between the real tACS and sham tACS for the PLVs, $p=0.699\pm0.189$.
      \par
\begin{figure}[h!]
 \begin{center}
   \includegraphics[width=13cm]{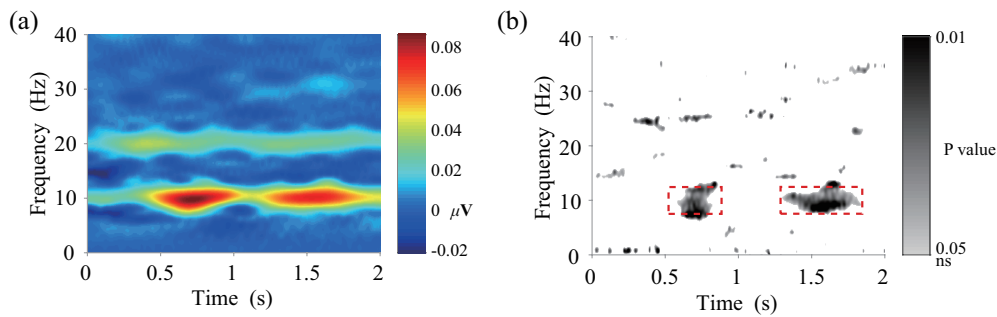}
   \caption{Change in time-frequency representation following tACS and its associated statistical significance. Values in the left panel were calculated by subtracting the change in time-frequency representation (post minus pre) in sham tACS from that of real tACS. The right panel illustrates the statistical significance, where p values greater than 0.05 are colored white and darker shades indicate greater significance. Red dashed rectangles outline the statistical significance corresponding to the left panel. EEG epochs on Oz electrode in baseline and post-1 blocks were analyzed.}    
   \label{fig:4}  
 \end{center}
\end{figure}

 As can be seen from Fig.\ref{fig:4}a, a remarkable increase was observed in the low frequencies of the spectrogram when contrasting real tACS with sham tACS. Specifically, an evident boost was shown in 10-Hz power, i.e., the fundamental frequency of SSVEP, along with the power of the second harmonic, 20Hz. Interestingly, the vicinity of the fundamental and second harmonic also exhibited an increase in power following tACS, with a bandwidth of approximate 4 Hz. Fig.\ref{fig:4}b illustrated the statistical significance corresponding to Fig.\ref{fig:4}a. Despite there being scattered gray dots in the statistical map (may be false positive), two gray shaded regions were formulated, indicating powers were enhanced significantly from 7 Hz to 13 Hz during 0.6 to 0.8 s, and also from 8 Hz to 13 Hz during 1.3 to 1.8 s. It was noteworthy that the statistically significant regions were centered at approximately 10 Hz. Additionally, the changes in third or fourth harmonic or even higher harmonics (not shown in Fig.\ref{fig:4}a) were not noticeable in the result.
     \par
\begin{figure}[h!]
 \begin{center}
   \includegraphics[width=13cm]{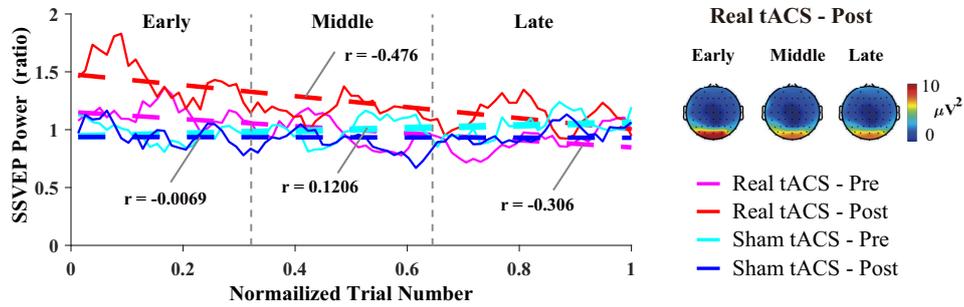}
   \caption{Temporal progression of single-trial SSVEP power for both conditions in baseline (Pre) and post-1 session (Post). The abscissa denotes normalized trial number within a block, i.e., trial numbers from 1 to 78 were normalized to 0 to 1 (early stage: 1$\sim$26 trials; middle stage: 27$\sim$52 trials; late stage: 53$\sim$78 trials). The thick dashed line indicates the linear regression of SSVEP power series with its associated regression coefficients on display. The right panel illustrates SSVEP spatial patterns during early, middle and late stages in real tACS-Post. Note that SSVEP powers were highest in the early stage and then were on a gradual decline for real tACS-Post.}    
   \label{fig:5}  
 \end{center}
\end{figure}

 In terms of the temporal progression of single-trial SSVEP power, the two conditions of real tACS and sham tACS showed distinct dynamic properties, as illustrated in Fig.\ref{fig:5}. The abscissa represented the normalized trial number, where 1/3 denoted the 26th trial, 2/3 denoted the 52nd trial, and so forth. Strikingly, at the early stage of the block, i.e., $0\sim1/3$ ($1\sim26$) trials, the SSVEP power in real tACS was significantly greater in the post-session ($10.636\pm1.519$ $\mu V^2$) than its baseline session ($8.324\pm1.221$ $\mu V^2$), $p=0.0131$, whereas in sham tACS no significant difference was found in post-session ($7.080\pm1.374$ $\mu V^2$) versus its baseline session ($7.169\pm1.007$ $\mu V^2$), $p=0.9350$. For the middle stage of the block, i.e., $1/3\sim2/3$ ($27\sim52$) trials, the mean SSVEP power was also greater in post-session ($8.543\pm1.225$ $\mu V^2$) than the baseline session ($7.119\pm0.905$ $\mu V^2$) of real tACS, though no significant differences were found in real tACS ($p=0.109$) and also in sham tACS ($p=0.286$; post:$6.561\pm0.868$ $\mu V^2$; baseline:$7.401\pm1.147$ $\mu V^2$). For the late stage of the block, i.e., $2/3\sim1$ ($53\sim78$) trials, in real tACS significant difference (p=0.0428) was found between post-session ($8.230\pm1.058$ $\mu V^2$) and baseline session ($7.238\pm0.976$ $\mu V^2$), whereas there was no significant difference in the sham tACS (post:$7.238\pm0.976$ $\mu V^2$; baseline:$7.764\pm1.136$ $\mu V^2$). We therefore noted that SSVEP power was the highest in the initial stage for the post block of real tACS. In other words, the intervention of tACS heightened the level of SSVEP power immediately after the end of stimulation, and then the aftereffect gradually tended to vanish, as indicated by the dashed line of linear regression. The regression coefficient was steeper in the post block of real tACS ($r=-0.476$) than the baseline block of real tACS ($r=-0.306$). Yet, the change for the sham tACS condition showed a tendency of slight increase following tACS (post: $r=-0.0069$; baseline: $r=-0.1206$). The spatial patterns of the post block (Fig.\ref{fig:5} right) in real tACS also confirmed the result of temporal progression, as was evident from the remarkably intense distribution of SSVEP during the early stage and its decline in the middle and late stage.
     \par
\begin{figure}[h!]
 \begin{center}
   \includegraphics[width=15cm]{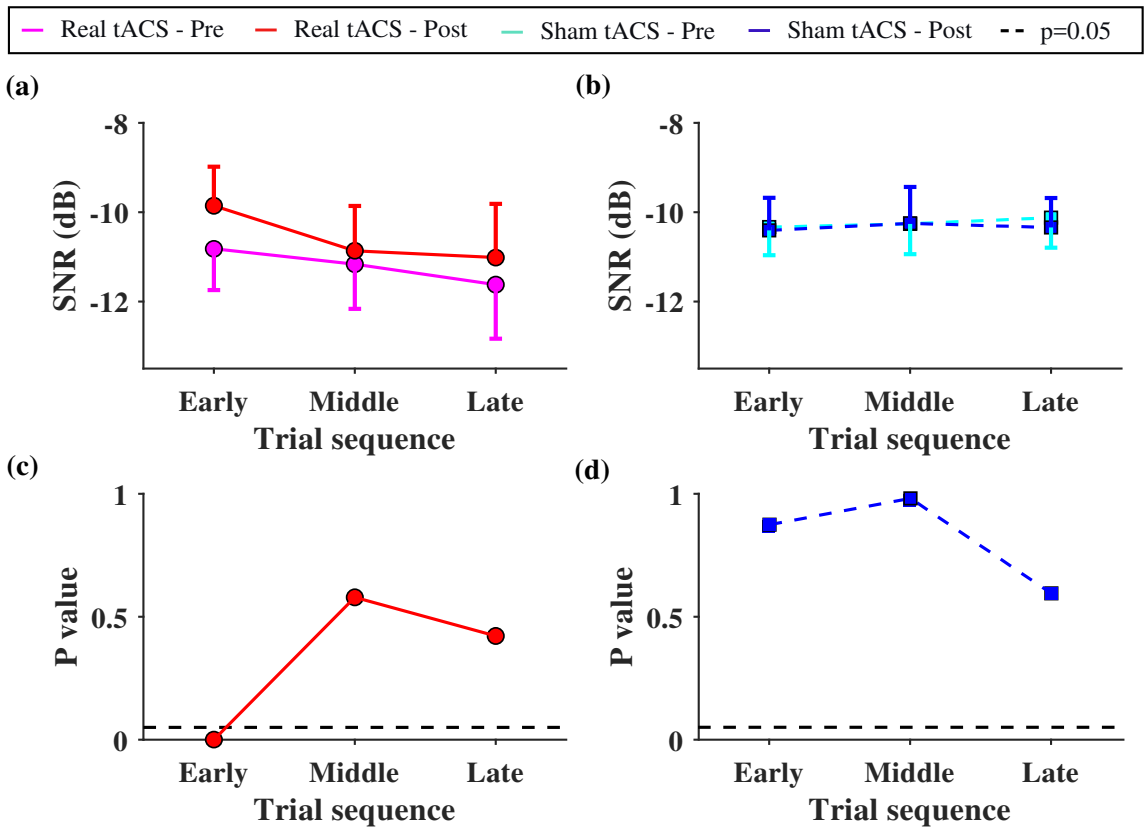}
   \caption{Change in SNR from baseline to the post-1 session and its statistical significance. The SNR was averaged within the early stage (1$\sim$26 trials), middle stage (27$\sim$52 trials) and late stage of a block (53$\sim$78 trials), respectively. A planned paired t-test calculates the statistical significance between the post-1 session and its baseline. The dashed line indicates a p-value of 0.05. (a). Comparison of SNR values for real tACS. (b). Comparison of SNR values for sham tACS. (c). Change in the statistical significance of SNR values for real tACS. (d). Change in the statistical significance of SNR values for sham tACS.}    
   \label{fig:6}  
 \end{center}
\end{figure}

 Fig.\ref{fig:6} illustrated the change in mean SNR during the three stages (early, middle and late) for real tACS (a) and sham tACS (b). The corresponding statistical significance between SNR in baseline and the post-1 session was illustrated in the bottom panel (c: real-tACS; d:sham tACS). As assessed by the Shapiro-Wilk test, the SNR data were normally distributed ($p<0.05$) for real tACS ($p=0.467\pm0.133$) and sham tACS ($p=0.636\pm0.151$). For real tACS, the SNR in the post-1 session was significantly greater than its baseline in the early stage ($p=0.0056$; post-1:$-9.856\pm0.873$dB; baseline:$-10.820\pm0.923$dB). Similar to the single-trial SSVEP power, a tendency of decay in SNR can be observed. For the middle and late stage, the differences between SNR in post-1 session and baseline were not significant (middle: $p=0.5786$; late: $p=0.4215$). For sham tACS, no significant changes in SNR from baseline to the post-1 session could be found in the early ($p=0.8745$), middle ($p=0.9816$) and late stage ($p=0.5946$).
     \par To ensure the boosted SSVEP power and SNR were independent of the change in spontaneous alpha power, we calculated the relative alpha power and relative 10-Hz power in the rest EEG. The Shapiro-Wilk test revealed that the assumption of normality was met for the relative alpha power ($p<0.05$, $p=0.344\pm0.078$) and the relative 10-Hz power ($p<0.05$, $p=0.562\pm0.097$). Mauchly's test of sphericity indicated that the assumption of sphericity was met for the two-way interaction, $\chi=15.741$, $p=0.076$ for relative alpha power and $\chi=10.97$, $p=0.285$ for the relative 10-Hz power. There was no significant two-way interaction between tACS and time for the relative alpha power, $F(4,44)=2.495$, $p=0.056$ and for relative 10-Hz power, $F(4,44)=1.619$, $p=0.186$. For relative alpha power, no significant main effect of tACS ($F(1,11)=0.642$, $p=0.44$) or time ($F(4,44)=0.299$, $p=0.877$) was found. For relative 10-Hz power, there was also no significant main effect of tACS ($F(1,11)=0.974$, $p=0.345$) or time ($F(4,44)=0.280$, $p=0.899$).
 \subsection{Behavioral data}     
      We further analyzed the VAS score to compare the participants' actual sensations in fatigue and discomfort under the two conditions. As expected, no significant difference existed in the VAS scores of discomfort evaluation between real tACS and sham tACS, $p=0.753$. With regard to the fatigue evaluation, two-way repeated measures ANOVA revealed that no significant interaction between tACS and time was found, $F(2.761, 30.373)=0.428$, $p=0.827$, $\epsilon=0.552$. The main effect of time showed a statistically significant difference in VAS of fatigue between blocks, $F(1.882, 20.703)=5.350$, $p=0.015$, $\epsilon=0.376$. Post hoc comparisons with a Bonferroni adjustment revealed that the significance existed in post-1 versus post-4, $p=0.040$, and post-2 versus post-3, $p=0.028$. However, there was no statistically significant effect of tACS on VAS of fatigue between real tACS and sham tACS for each block, $p=0.678\pm0.28$. In other terms, as time progressed to the middle period of the experiment, subjects tended to be more fatigue. But the administration of tACS condition was imperceptible to the subjects and it did not further introduce confounds such as sensations of fatigue or discomfort, physically or psychologically. Specifically, the extent of VAS on discomfort ($1.423\pm0.643$) was within the range of slight discomfort ($0\sim4$). For fatigue, VAS evaluations ($7.506\pm1.560$) were within the range of slight or moderate fatigue ($5\sim10$).
      \par In regard to the Go/No-Go task, four subjects failed to remember the target stimuli in some blocks and accordingly were removed for analysis. There was no significant two-way interaction for the d prime data, $F(5, 35)=1.519$, $p=0.209$, and no main effect of tACS ($p=0.750$) or time ($p=0.541$) on d prime. Also for the RT data, no significant interaction was found, $F(5,35)=1.302$, $p=0.286$. The main effect of tACS or time showed no statistical significance in reaction time, $p=0.993$ and $p=0.096$, respectively.

\section{Discussion}
      In this paper, we evaluated the effects of tACS on SSVEP both at 10 Hz in a group of healthy subjects and demonstrated the feasibility of modulating SSVEP using tACS. The current findings provide direct evidence for boosting SSVEP by a manual intervention of tACS. Since SSVEP is an indication of cortical excitability\cite{schomer2012niedermeyer}, this study lends support for the utility of tACS previously reported in altering the excitability of human cortex\cite{kanai2008frequency-dependent, romei2008spontaneous}. Overall, the present study yielded three-fold findings. First, the administration of 10-Hz tACS significantly enhanced the 10-Hz SSVEP, as was validated by the spectral, temporal, spatial and SNR profiles of SSVEP. Second, the aftereffect of tACS on SSVEP achieved climax immediately after the termination of tACS, and gradually receded throughout trials and blocks. Third, the application of tACS did not affect the behavioral performance during the SSVEP task, as assessed by d prime and reaction time when subjects underwent the Go/No-Go task. 
      In the following paragraphs, we discuss some issues in detail concerning the aftereffect and the study.
     \par  Most existing tACS studies adopted the approach of a pre-post stimulation comparison and consequently were unable to analyze the maximal duration of tACS aftereffect\cite{herrmann2017can}. In this study, we tackled this issue and intended to determine the span of the aftereffect of tACS. The experimental design of a single flickering condition in our study ensured a decent amount of trials obtained from one block. In this fashion, the repeated measures made it possible to gauge the dynamic change of SSVEPs with time course. As such, the findings in Fig.\ref{fig:1} and Fig.\ref{fig:2} indicate that the significant excitatory effect on SSVEP persisted for merely the post-1 block, i.e., a period of 5 minutes following tACS. The short duration of tACS aftereffect is in line with the previous studies (5 min\cite{bergmann2009acute, schutter2011brain}, 3 min\cite{struber2014antiphasic}, 1 min\cite{helfrich2014selective}) and also resembles the 5-minute aftereffect by transcranial magnetic stimulation (TMS)\cite{veniero2011alpha-generation}, considering the physiological effects of TMS and transcranial electrical stimulation (TES) are comparable\cite{brocke2005transcranial}. On closer examination within the post-1 block, the initial upsurge and subsequent trend of decline in single-trial SSVEP power characterize the lingering and reversible effect of tACS. Empirically, the duration of aftereffect in tACS studies depend mainly on the duration\cite{ neuling2013orchestrating, kasten2016sustained, struber2015on}, montage\cite{ruhnau2016flicker-driven, mehta2015montage}, intensity and frequency of stimulation\cite{antal2013transcranial} and the brain state\cite{neuling2013orchestrating}.  We employed the parameters of tACS in line with previous studies (intensity\cite{neuling2015friends}, duration\cite{neuling2013orchestrating} and montage\cite{kasten2016sustained}), thus the short-lasting aftereffect in our findings may result from the task-dependent feature of tACS aftereffect\cite{cabralcalderin2016transcranial}. Previous research in the domain of transcranial direct current stimulation (tDCS) found that the duration of aftereffect induced by visual cortical tDCS is relatively short compared to the aftereffects induced by motor cortical tDCS\cite{antal2008transcranial, antal2011electrical}. Speculatively, the primary visual cortex may be less tunable than the primary motor cortex because of different neuronal membrane properties and cortical connections influencing neuroplasticity\cite{antal2008transcranial, antal2011electrical}. This may provide us some insights since visual evoked potentials (VEPs) elicited by visual tasks are not well known under the intervention of tACS. It also cannot exclude the possibility that the aftereffect on cortical excitability of tACS was more reversible and susceptible to be counterbalanced by the predominantly provoked steady-state responses in the primary visual cortex. 
      \par Since SSVEP and tACS are both characterized by frequency modulation, the interplay between these two rhythmic endogenous and exogenous signals are attractive and intriguing to the researchers. However, to our knowledge, few studies have been involved in this domain. A recent study\cite{ruhnau2016flicker-driven} investigated the online effect of tACS in a magnetoencephalographic (MEG) setting on 7-Hz and 11-Hz visual SSRs (steady-state responses), a.k.a. SSVEPs. It was discovered that same-frequency tACS did not significantly affect the fundamental frequency but significantly enhanced the higher harmonics of SSVEP, which was not consistent with the result of our study. As a matter of fact, both studies utilized the tACS and SSVEP in the alpha range (8-12 Hz), i.e., 10 Hz (ours) and 11 Hz (the former), and the stimulation intensities are comparable (ours: 650$\mu$V, the former: 613$\pm$128$\mu$V). As suggested by the verified model of the effect of tACS on spontaneous EEG\cite{merlet2013from}, the frequency of 10 Hz and 11 Hz might share common tuning effects, though nuance in intensity. Therefore, the distinction in frequency might not suffice to account for the disparity of result. As such, on the one hand, we should notice that the former study employed 2-s tACS stimulation trial by trial. This very short intermittent protocol has been demonstrated too short in stimulation duration to induce tACS aftereffect\cite{struber2015on, vossen2015alpha}. Thus the long span of 20-min stimulation in our study might offer an advantage to probe into the tACS aftereffect on SSVEP. On the other hand, in the former study, considerable tACS harmonics still existed in the reconstructed source space, and the author admitted the after-effects were out of the scope of the study. In this view, the present study serves as an improvement to the former study and provide the first informative evidence to unveil the aftereffect.
      
      \par The pronounced SSVEP in our study is indicative of elevated excitability of visual cortex, which is consistent with other tACS-visual studies demonstrating tACS enhanced visual cortical excitability as measured by phosphenes\cite{kanai2008frequency-dependent, kanai2010transcranial}. Note that as expected, all subjects in our study did not report phosphene perception during the administration of tACS due to the choice of stimulation intensity\cite{neuling2015friends}. Apart from the elevated SSVEP in the occipital region, a noticeable statistical difference can also be observed in the left temporal and parietal region in the post-1 and the right frontal region in the post-4 (Fig.\ref{fig:1}). Since our study utilizes a central flicker that was different from the stimuli with eccentricity \cite{maye2017utilizing}, few SSVEP is evoked in the regions and the values may represent propagation of SSVEP from the occipital region \cite{thorpe2007identification} or background EEG during the visual task. It is interesting to note that following tACS a marked increase in power was likewise elevated in the spectral vicinity of SSVEP harmonics, which may result from the facilitation of photic driving \cite{sakamoto1993preservation} by tACS. For the finding of rest EEG, the present study reveals no significant influence of tACS on the spontaneous alpha power following tACS. The negative result on spontaneous alpha is reported in recent studies \cite{clayton2018effects,stecher2018absence} and may be ascribed to the difference in stimulation frequency which the majority of previous studies \cite{zaehle2010transcranial,neuling2013orchestrating,vossen2015alpha,kasten2016sustained,stecher2017ten} employ at an individual alpha frequency (IAF) of each subject. In the context of the present study, the absence of tACS effect on spontaneous alpha further supports the notion of an increase in SNR of SSVEP in our finding. For the behavior findings, the result revealed that no modulatory effect of tACS was found on sensitivity index and reaction time in the Go/No-Go task. This negative report is in line with a previous study \cite{brauer2018no} of Go/No-Go task and other attention studies using 10-Hz tACS\cite{wittenberg201910,clayton2018effects,van2018no,sheldon2018does}. In addition, the comparable VAS ratings between real and sham tACS indicate the effective control that rules out indirect confounds such as physiological or psychological sensations.  
      \par As for the neurophysiological underpinnings of tACS aftereffect, there has been a growing debate about whether the mechanisms of neuronal entrainment\cite{thut2011entrainment, helfrich2014entrainment, romei2016information-based} or neural plasticity\cite{vossen2015alpha, zaehle2010transcranial, veniero2015lasting} actually play the casual role. 
      From the evidence of sinusoidal attributes of tACS and SSVEP alone, it is plausible to lend support for the theory of neuronal entrainment. As is indicated by a previous study \cite{herrmann2001human}, SSVEP shows resonance frequencies, i.e., strong resonance peaks around 10 Hz and weak peaks around 20 and 40 Hz, etc. across 1--100Hz, a phenomenon which is in accord with the entrainment of ongoing natural oscillation and conforms to the theory\cite{thut2011entrainment}. The results in the present study (Fig.\ref{fig:4}a) clearly showcases a significant increase in 10-Hz SSVEP power and also slight up-regulation in 20-Hz SSVEP power, which may imply the resonance attribute of entrainment in the modulation of SSVEP by tACS at first glance. Nonetheless, two pieces of evidence in the present study provide an argument against the hypothesis of entrainment. First, the transcranial electrical stimulation both with and without sinusoidal attribute yield similar modulatory result for SSVEP. The finding in our previous study\cite{liu2017effects} on tDCS manifested facilitatory effects on 10-Hz SSVEP in a similar fashion though statistical significant at a smaller scale. This to some extent implies that the aftereffects in the two studies may have in common underlying mechanism that is independent of neuronal entrainment, albeit the epiphenomenon of the brain might slightly vary. Second, the phase-locking values in Fig.\ref{fig:3}b, as a measurement of synchronization between the sinusoidal visual stimuli and SSVEPs, would have been perturbed by the aftereffect of entrainment if the hypothesis holds\cite{vossen2015alpha,haberbosch2019rebound}. However, no significant changes were found in the phase-locking values between real tACS and sham tACS after the intervention. Taken together, the findings in our study are not consistent with the hypothesis of entrainment and therefore provide evidence in favor of the neural plasticity theory.
      \par From the perspective of tACS application, the finding in this study opens a new avenue for boosting SSVEP-BCI performance by transcranially heightening the SNR of visual responses. Since SNR is correlated with information transfer rate (ITR) in BCI classification, the increment in SNR of SSVEP implies the possibility of enhancing the ITR and aiding in BCI training via tACS neuromodulation. A previous pilot study reported a positive effect of tACS (1mA; 10Hz; 10min; PO9-PO10 protocol) on six subjects who performed a three-target (12.5Hz, 9.37Hz, and 8.33Hz) BCI task \cite{duan2016effects}. The result of our study is in line with the previous study and provides the underpinning for the integration of tACS into SSVEP-BCI task. At a population level, the beneficial effect of tACS on SNR also proposes potential solutions for individuals of low SNR and ITR, i.e., BCI illiteracy. Nevertheless, the significant improvement in SNR of SSVEP carries on at a short time-scale, i.e., the early stage of the post-1 session in the present study. This indicates the short duration of the aftereffect on SNR and also the reversible attribute of tACS effect that is favorable from the standpoint of real-world application. For future BCI studies, more reinforced measures, e.g. repeated sessions \cite{schmidt2013progressive}, high-definition montage \cite {berger2018brain} could be taken for the prolongation of the lingering effect. 

\section{Conclusion}
       To summarize, our study intends to explore the possibility of brain state manipulation by combing visual stimulation and rhythmic stimulation\cite{romei2016information-based}. Specifically, by applying strict protocols of 10-Hz tACS on 10-Hz SSVEP, the current study measures the longitudinal changes of SSVEP between blocks and within a block and provides the first evidence that tACS boosts SSVEP in a short period of time. Importantly, this indicates that tACS strengthens brain excitability and facilitates flickering driving and may serve as a novel tool of neural feedback, which may provide new insight for basic neuroscience and unlock new applications. 
\section{Acknowledgment}
       We would like to thank faculty and students at Neural Engineering at Tsinghua University for their helpful discussion. This research was supported by the National Natural Science Foundation of China under grant 61431007, 61603416, 91220301, 91320202 and the National Key R$\&$D Program of China under grant 2017YFB1002505.
\section{References}

%


\begin{thebibliography}{89}
\expandafter\ifx\csname url\endcsname\relax
  \def\url#1{\texttt{#1}}\fi
\expandafter\ifx\csname urlprefix\endcsname\relax\def\urlprefix{URL }\fi
\expandafter\ifx\csname href\endcsname\relax
  \def\href#1#2{#2} \def\path#1{#1}\fi

\bibitem{yavari2017basic}
F.~Yavari, A.~Jamil, M.~M. Samani, L.~P. Vidor, M.~A. Nitsche, Basic and
  functional effects of transcranial electrical stimulation (tes)—an
  introduction, Neuroscience \& Biobehavioral Reviews.

\bibitem{kanai2008frequency-dependent}
R.~Kanai, L.~Chaieb, A.~Antal, V.~Walsh, W.~Paulus, Frequency-dependent
  electrical stimulation of the visual cortex, Current Biology 18~(23) (2008)
  1839--1843.

\bibitem{zaehle2010transcranial}
T.~Zaehle, S.~Rach, C.~S. Herrmann, Transcranial alternating current
  stimulation enhances individual alpha activity in human eeg, PLOS ONE 5~(11).

\bibitem{merlet2013from}
I.~Merlet, G.~Birot, R.~Salvador, B.~Molaeeardekani, A.~Mekonnen,
  A.~Soriafrish, G.~Ruffini, P.~C. Miranda, F.~Wendling, From oscillatory
  transcranial current stimulation to scalp eeg changes: A biophysical and
  physiological modeling study, PLOS ONE 8~(2).

\bibitem{neuling2013orchestrating}
T.~Neuling, S.~Rach, C.~S. Herrmann, Orchestrating neuronal networks: sustained
  after-effects of transcranial alternating current stimulation depend upon
  brain states., Frontiers in Human Neuroscience 7 (2013) 161--161.

\bibitem{helfrich2014entrainment}
R.~F. Helfrich, T.~R. Schneider, S.~Rach, S.~A. Trautmannlengsfeld, A.~Engel,
  C.~S. Herrmann, Entrainment of brain oscillations by transcranial alternating
  current stimulation., Current Biology 24~(3) (2014) 333--339.

\bibitem{vosskuhl2016bold}
J.~Vosskuhl, R.~J. Huster, C.~S. Herrmann, Bold signal effects of transcranial
  alternating current stimulation (tacs) in the alpha range: A concurrent
  tacs–fmri study, NeuroImage 140 (2016) 118--125.

\bibitem{cabral2016transcranial}
Y.~Cabral-Calderin, K.~A. Williams, A.~Opitz, P.~Dechent, M.~Wilke,
  Transcranial alternating current stimulation modulates spontaneous low
  frequency fluctuations as measured with fmri, Neuroimage 141 (2016) 88--107.

\bibitem{bergmann2009acute}
T.~O. Bergmann, S.~Groppa, M.~Seeger, M.~Molle, L.~Marshall, H.~R. Siebner,
  Acute changes in motor cortical excitability during slow oscillatory and
  constant anodal transcranial direct current stimulation., Journal of
  Neurophysiology 102~(4) (2009) 2303--2311.

\bibitem{struber2014antiphasic}
D.~Struber, S.~Rach, S.~A. Trautmannlengsfeld, A.~Engel, C.~S. Herrmann,
  Antiphasic 40 hz oscillatory current stimulation affects bistable motion
  perception, Brain Topography 27~(1) (2014) 158--171.

\bibitem{helfrich2014selective}
R.~F. Helfrich, H.~Knepper, G.~Nolte, D.~Struber, S.~Rach, C.~S. Herrmann,
  T.~R. Schneider, A.~Engel, Selective modulation of interhemispheric
  functional connectivity by hd-tacs shapes perception., PLOS Biology 12~(12).

\bibitem{kasten2016sustained}
F.~H. Kasten, J.~Dowsett, C.~S. Herrmann, Sustained aftereffect of α-tacs
  lasts up to 70 min after stimulation., Frontiers in Human Neuroscience 10
  (2016) 245.

\bibitem{ruhnau2016eyes}
P.~Ruhnau, T.~Neuling, M.~Fusca, C.~S. Herrmann, G.~Demarchi, N.~Weisz, Eyes
  wide shut: Transcranial alternating current stimulation drives alpha rhythm
  in a state dependent manner., Scientific Reports 6~(1) (2016) 27138.

\bibitem{antal2008comparatively}
A.~Antal, K.~Boros, C.~Poreisz, L.~Chaieb, D.~Terney, W.~Paulus, Comparatively
  weak after-effects of transcranial alternating current stimulation (tacs) on
  cortical excitability in humans, Brain stimulation 1~(2) (2008) 97--105.

\bibitem{pogosyan2009boosting}
A.~Pogosyan, L.~D. Gaynor, A.~Eusebio, P.~Brown, Boosting cortical activity at
  beta-band frequencies slows movement in humans, Current biology 19~(19)
  (2009) 1637--1641.

\bibitem{joundi2012driving}
R.~A. Joundi, N.~Jenkinson, J.-S. Brittain, T.~Z. Aziz, P.~Brown, Driving
  oscillatory activity in the human cortex enhances motor performance, Current
  Biology 22~(5) (2012) 403--407.

\bibitem{reinhart2019working}
R.~M. Reinhart, J.~A. Nguyen, Working memory revived in older adults by
  synchronizing rhythmic brain circuits, Nature neuroscience (2019) 1.

\bibitem{marshall2006boosting}
L.~Marshall, H.~Helgad{\'o}ttir, M.~M{\"o}lle, J.~Born, Boosting slow
  oscillations during sleep potentiates memory, Nature 444~(7119) (2006)
  610--613.

\bibitem{kirov2009slow}
R.~Kirov, C.~Weiss, H.~R. Siebner, J.~Born, L.~Marshall, Slow oscillation
  electrical brain stimulation during waking promotes eeg theta activity and
  memory encoding, Proceedings of the National Academy of Sciences 106~(36)
  (2009) 15460--15465.

\bibitem{santarnecchi2013frequency}
E.~Santarnecchi, N.~R. Polizzotto, M.~Godone, F.~Giovannelli, M.~Feurra,
  L.~Matzen, A.~Rossi, S.~Rossi, Frequency-dependent enhancement of fluid
  intelligence induced by transcranial oscillatory potentials, Current Biology
  23~(15) (2013) 1449--1453.

\bibitem{uhlhaas2012neuronal}
P.~J. Uhlhaas, W.~Singer, Neuronal dynamics and neuropsychiatric disorders:
  toward a translational paradigm for dysfunctional large-scale networks,
  Neuron 75~(6) (2012) 963--980.

\bibitem{brittain2013tremor}
J.-S. Brittain, P.~Probert-Smith, T.~Z. Aziz, P.~Brown, Tremor suppression by
  rhythmic transcranial current stimulation, Current Biology 23~(5) (2013)
  436--440.

\bibitem{ahn2019targeting}
S.~Ahn, J.~M. Mellin, S.~Alagapan, M.~L. Alexander, J.~H. Gilmore, L.~F.
  Jarskog, F.~Frohlich, Targeting reduced neural oscillations in patients with
  schizophrenia by transcranial alternating current stimulation, NeuroImage 186
  (2019) 126--136.

\bibitem{laczo2012transcranial}
B.~Laczo, A.~Antal, R.~Niebergall, S.~Treue, W.~Paulus, Transcranial
  alternating stimulation in a high gamma frequency range applied over v1
  improves contrast perception but does not modulate spatial attention, Brain
  Stimulation 5~(4) (2012) 484--491.

\bibitem{schutter2016cutaneous}
D.~J. Schutter, Cutaneous retinal activation and neural entrainment in
  transcranial alternating current stimulation: a systematic review, Neuroimage
  140 (2016) 83--88.

\bibitem{brignani2013transcranial}
D.~Brignani, M.~Ruzzoli, P.~Mauri, C.~Miniussi, Is transcranial alternating
  current stimulation effective in modulating brain oscillations?, PloS one
  8~(2) (2013) e56589.

\bibitem{herring2019low-frequency}
J.~D. Herring, S.~Esterer, T.~R. Marshall, O.~Jensen, T.~O. Bergmann,
  Low-frequency alternating current stimulation rhythmically suppresses
  gamma-band oscillations and impairs perceptual performance, NeuroImage 184
  (2019) 440--449.

\bibitem{ruhnau2016flicker-driven}
P.~Ruhnau, C.~Keitel, C.~Lithari, N.~Weisz, T.~Neuling, Flicker-driven
  responses in visual cortex change during matched-frequency transcranial
  alternating current stimulation, Frontiers in Human Neuroscience 10 (2016)
  184.

\bibitem{morgan1996selective}
S.~T. Morgan, J.~C. Hansen, S.~A. Hillyard, Selective attention to stimulus
  location modulates the steady-state visual evoked potential, Proceedings of
  the National Academy of Sciences of the United States of America 93~(10)
  (1996) 4770--4774.

\bibitem{muller1998the}
M.~M. Muller, W.~A. Tedersalejarvi, S.~A. Hillyard, The time course of cortical
  facilitation during cued shifts of spatial attention., Nature Neuroscience
  1~(7) (1998) 631--634.

\bibitem{muller2003sustained}
M.~M. Muller, P.~Malinowski, T.~Gruber, S.~A. Hillyard, Sustained division of
  the attentional spotlight., Nature 424~(6946) (2003) 309--312.

\bibitem{schomer2012niedermeyer}
D.~L. Schomer, F.~L. Da~Silva, Niedermeyer's electroencephalography: basic
  principles, clinical applications, and related fields, Lippincott Williams \&
  Wilkins, 2012.

\bibitem{medeiros2016ngoggle}
F.~A. Medeiros, J.~K. Zao, Y.~Wang, M.~Nakanishi, Y.-P. Lin, A.~Diniz-Filho,
  T.-P. Jung, The ngoggle: a portable brain-based method for assessment of
  visual function deficits in glaucoma, Investigative Ophthalmology \& Visual
  Science 57~(12) (2016) 3940--3940.

\bibitem{regan1989human}
D.~Regan, Human brain electrophysiology: evoked potentials and evoked magnetic
  fields in science and medicine, Elsevier, 1989.

\bibitem{herrmann2001human}
C.~S. Herrmann, Human eeg responses to 1-100 hz flicker: resonance phenomena in
  visual cortex and their potential correlation to cognitive phenomena,
  Experimental Brain Research 137 (2001) 346--353.

\bibitem{vialatte2010steady-state}
F.~Vialatte, M.~Maurice, J.~Dauwels, A.~Cichocki, Steady-state visually evoked
  potentials: focus on essential paradigms and future perspectives., Progress
  in Neurobiology 90~(4) (2010) 418--438.

\bibitem{gao2003a}
X.~Gao, D.~Xu, M.~Cheng, S.~Gao, A bci-based environmental controller for the
  motion-disabled, international conference of the ieee engineering in medicine
  and biology society 11~(2) (2003) 137--140.

\bibitem{chen2015high-speed}
X.~Chen, Y.~Wang, M.~Nakanishi, X.~Gao, T.~Jung, S.~Gao, High-speed spelling
  with a noninvasive brain-computer interface., Proceedings of the National
  Academy of Sciences of the United States of America 112~(44).

\bibitem{ozen2010transcranial}
S.~Ozen, A.~Sirota, M.~Belluscio, C.~A. Anastassiou, E.~Stark, C.~Koch,
  G.~Buzsaki, Transcranial electric stimulation entrains cortical neuronal
  populations in rats, The Journal of Neuroscience 30~(34) (2010) 11476--11485.

\bibitem{notbohm2016modification}
A.~Notbohm, J.~Kurths, C.~S. Herrmann, Modification of brain oscillations via
  rhythmic light stimulation provides evidence for entrainment but not for
  superposition of event-related responses, Frontiers in Human Neuroscience 10
  (2016) 10.

\bibitem{capilla2011steady-state}
A.~Capilla, P.~Pazoalvarez, A.~Darriba, P.~Campo, J.~Gross, Steady-state visual
  evoked potentials can be explained by temporal superposition of transient
  event-related responses., PLOS ONE 6~(1).

\bibitem{herrmann2013transcranial}
C.~S. Herrmann, S.~Rach, T.~Neuling, D.~Struber, Transcranial alternating
  current stimulation: a review of the underlying mechanisms and modulation of
  cognitive processes, Frontiers in Human Neuroscience 7 (2013) 279--279.

\bibitem{brainard1997psychophysics}
D.~H. Brainard, S.~Vision, The psychophysics toolbox, Spatial vision 10 (1997)
  433--436.

\bibitem{toscani2010alpha}
M.~Toscani, T.~Marzi, S.~Righi, M.~P. Viggiano, S.~Baldassi, Alpha waves: a
  neural signature of visual suppression, Experimental Brain Research 207
  (2010) 213--219.

\bibitem{neuling2012finite-element}
T.~Neuling, S.~Wagner, C.~H. Wolters, T.~Zaehle, C.~S. Herrmann, Finite-element
  model predicts current density distribution for clinical applications of tdcs
  and tacs, Frontiers in Psychiatry 3 (2012) 83--83.

\bibitem{neuling2015friends}
T.~Neuling, P.~Ruhnau, M.~Fusca, G.~Demarchi, C.~S. Herrmann, N.~Weisz,
  Friends, not foes: Magnetoencephalography as a tool to uncover brain dynamics
  during transcranial alternating current stimulation, NeuroImage 118 (2015)
  406--413.

\bibitem{nosek2001go}
B.~A. Nosek, M.~R. Banaji, The go/no-go association task, Social cognition
  19~(6) (2001) 625--666.

\bibitem{knotkova2016textbook}
H.~Knotkova, D.~Rasche, et~al., Textbook of Neuromodulation, Springer, 2016.

\bibitem{wieser2014fearful}
M.~J. Wieser, A.~Keil, Fearful faces heighten the cortical representation of
  contextual threat., NeuroImage 86 (2014) 317--325.

\bibitem{silberstein1990steady-state}
R.~B. Silberstein, M.~Schier, A.~Pipingas, J.~Ciorciari, S.~R. Wood, D.~G.
  Simpson, Steady-state visually evoked potential topography associated with a
  visual vigilance task, Brain Topography 3~(2) (1990) 337--347.

\bibitem{silberstein2000steady-state}
R.~B. Silberstein, P.~Line, A.~Pipingas, D.~L. Copolov, P.~G. Harris,
  Steady-state visually evoked potential topography during the continuous
  performance task in normal controls and schizophrenia, Clinical
  Neurophysiology 111~(5) (2000) 850--857.

\bibitem{zhang2018a}
S.~Zhang, X.~Han, X.~Chen, Y.~Wang, S.~Gao, X.~Gao, A study on dynamic model of
  steady-state visual evoked potentials, Journal of Neural Engineering 15~(4)
  (2018) 046010.

\bibitem{lachaux1999measuring}
J.~Lachaux, E.~Rodriguez, J.~Martinerie, F.~J. Varela, Measuring phase
  synchrony in brain signals, Human Brain Mapping 8~(4) (1999) 194--208.

\bibitem{aydore2013a}
S.~Aydore, D.~A. Pantazis, R.~M. Leahy, A note on the phase locking value and
  its properties, NeuroImage 74 (2013) 231--244.

\bibitem{manyakov2013sampled}
N.~V. Manyakov, N.~Chumerin, A.~Robben, A.~Combaz, M.~Van~Vliet, M.~M.
  Van~Hulle, Sampled sinusoidal stimulation profile and multichannel fuzzy
  logic classification for monitor-based phase-coded ssvep brain-computer
  interfacing, Journal of Neural Engineering 10~(3) (2013) 036011.

\bibitem{Chen2014A}
X.~Chen, Z.~Chen, S.~Gao, X.~Gao, A high-itr ssvep-based bci speller,
  Brain-Computer Interfaces 1~(3-4) (2014) 181--191.

\bibitem{norcia2015the}
A.~M. Norcia, L.~G. Appelbaum, J.~Ales, B.~R. Cottereau, B.~Rossion, The
  steady-state visual evoked potential in vision research: A review, Journal of
  Vision 15~(6) (2015) 4--4.

\bibitem{stanislaw1999calculation}
H.~Stanislaw, N.~Todorov, Calculation of signal detection theory measures,
  Behavior Research Methods Instruments Computers 31~(1) (1999) 137--149.

\bibitem{romei2008spontaneous}
V.~Romei, V.~Brodbeck, C.~M. Michel, A.~Amedi, A.~Pascualleone, G.~Thut,
  Spontaneous fluctuations in posterior α-band eeg activity reflect
  variability in excitability of human visual areas, Cerebral Cortex 18~(9)
  (2008) 2010--2018.

\bibitem{herrmann2017can}
C.~S. Herrmann, D.~Str{\"u}ber, What can transcranial alternating current
  stimulation tell us about brain oscillations?, Current Behavioral
  Neuroscience Reports 4~(2) (2017) 128--137.

\bibitem{schutter2011brain}
D.~J. L.~G. Schutter, R.~Hortensius, Brain oscillations and frequency-dependent
  modulation of cortical excitability, Brain Stimulation 4~(2) (2011) 97--103.

\bibitem{veniero2011alpha-generation}
D.~Veniero, D.~Brignani, G.~Thut, C.~Miniussi, Alpha-generation as basic
  response-signature to transcranial magnetic stimulation (tms) targeting the
  human resting motor cortex: A tms/eeg co-registration study, Psychophysiology
  48~(10) (2011) 1381--1389.

\bibitem{brocke2005transcranial}
J.~Brocke, K.~Irlbacher, B.~Hauptmann, M.~Voss, S.~Brandt, Transcranial
  magnetic and electrical stimulation compared: does tes activate intracortical
  neuronal circuits?, Clinical neurophysiology 116~(12) (2005) 2748--2756.

\bibitem{struber2015on}
D.~Struber, S.~Rach, T.~Neuling, C.~S. Herrmann, On the possible role of
  stimulation duration for after-effects of transcranial alternating current
  stimulation., Frontiers in Cellular Neuroscience 9 (2015) 311--311.

\bibitem{mehta2015montage}
A.~R. Mehta, A.~Pogosyan, P.~Brown, J.~Brittain, Montage matters: the influence
  of transcranial alternating current stimulation on human physiological
  tremor., Brain Stimulation 8~(2) (2015) 260--268.

\bibitem{antal2013transcranial}
A.~Antal, W.~Paulus, Transcranial alternating current stimulation (tacs),
  Frontiers in Human Neuroscience 7 (2013) 317--317.

\bibitem{cabralcalderin2016transcranial}
Y.~Cabralcalderin, C.~A. Weinrich, C.~Schmidtsamoa, E.~Poland, P.~Dechent,
  M.~Bahr, M.~Wilke, Transcranial alternating current stimulation affects the
  bold signal in a frequency and task-dependent manner, Human Brain Mapping
  37~(1) (2016) 94--121.

\bibitem{antal2008transcranial}
A.~Antal, W.~Paulus, Transcranial direct current stimulation and visual
  perception, Perception 37~(3) (2008) 367--374.

\bibitem{antal2011electrical}
A.~Antal, W.~Paulus, M.~A. Nitsche, Electrical stimulation and visual network
  plasticity, Restorative Neurology and Neuroscience 29~(6) (2011) 365--374.

\bibitem{vossen2015alpha}
A.~Vossen, J.~Gross, G.~Thut, Alpha power increase after transcranial
  alternating current stimulation at alpha frequency (α-tacs) reflects plastic
  changes rather than entrainment, Brain Stimulation 8~(3) (2015) 499--508.

\bibitem{kanai2010transcranial}
R.~Kanai, W.~Paulus, V.~Walsh, Transcranial alternating current stimulation
  (tacs) modulates cortical excitability as assessed by tms-induced phosphene
  thresholds, Clinical Neurophysiology 121~(9) (2010) 1551--1554.

\bibitem{maye2017utilizing}
A.~Maye, D.~Zhang, A.~K. Engel, Utilizing retinotopic mapping for a
  multi-target ssvep bci with a single flicker frequency, IEEE Transactions on
  Neural Systems and Rehabilitation Engineering 25~(7) (2017) 1026--1036.

\bibitem{thorpe2007identification}
S.~G. Thorpe, P.~L. Nunez, R.~Srinivasan, Identification of wave-like spatial
  structure in the ssvep: Comparison of simultaneous eeg and meg, Statistics in
  medicine 26~(21) (2007) 3911--3926.

\bibitem{sakamoto1993preservation}
H.~Sakamoto, T.~Inouye, K.~Shinosaki, Preservation of alpha rhythm shortly
  after photic driving, International Journal of Neuroscience 73 (1993)
  227--233.

\bibitem{clayton2018effects}
M.~S. Clayton, N.~Yeung, R.~Cohen~Kadosh, The effects of 10 hz transcranial
  alternating current stimulation on audiovisual task switching, Frontiers in
  neuroscience 12 (2018) 67.

\bibitem{stecher2018absence}
H.~I. Stecher, C.~Herrmann, Absence of alpha-tacs aftereffects in darkness
  reveals importance of taking derivations of stimulation frequency and
  individual alpha variability into account, Frontiers in Psychology 9.

\bibitem{stecher2017ten}
H.~I. Stecher, T.~M. Pollok, D.~Str{\"u}ber, F.~Sobotka, C.~S. Herrmann, Ten
  minutes of $\alpha$-tacs and ambient illumination independently modulate eeg
  $\alpha$-power, Frontiers in human neuroscience 11 (2017) 257.

\bibitem{brauer2018no}
H.~Brauer, N.~E. Kadish, A.~Pedersen, M.~Siniatchkin, V.~Moliadze, No
  modulatory effects when stimulating the right inferior frontal gyrus with
  continuous 6 hz tacs and trns on response inhibition: A behavioral study,
  Neural plasticity 2018.

\bibitem{wittenberg201910}
M.~A. Wittenberg, M.~Morr, A.~Schnitzler, J.~Lange, 10 hz tacs over
  somatosensory cortex does not modulate supra-threshold tactile temporal
  discrimination in humans., Frontiers in neuroscience 13 (2019) 311.

\bibitem{van2018no}
M.~R. Van~Schouwenburg, L.~K. Sorensen, R.~de~Klerk, L.~C. Reteig, H.~A.
  Slagter, No differential effects of two different alpha-band electrical
  stimulation protocols over fronto-parietal regions on spatial attention.,
  Frontiers in neuroscience 12 (2018) 433.

\bibitem{sheldon2018does}
S.~S. Sheldon, K.~E. Mathewson, Does 10-hz cathodal oscillating current of the
  parieto-occipital lobe modulate target detection?, Frontiers in Neuroscience
  12 (2018) 83.

\bibitem{thut2011entrainment}
G.~Thut, P.~G. Schyns, J.~Gross, Entrainment of perceptually relevant brain
  oscillations by non‑invasive rhythmic stimulation of the human brain,
  Frontiers in Psychology 2 (2011) 170--170.

\bibitem{romei2016information-based}
V.~Romei, G.~Thut, J.~Silvanto, Information-based approaches of noninvasive
  transcranial brain stimulation, Trends in Neurosciences 39~(11) (2016)
  782--795.

\bibitem{veniero2015lasting}
D.~Veniero, A.~Vossen, J.~Gross, G.~Thut, Lasting eeg/meg aftereffects of
  rhythmic transcranial brain stimulation: Level of control over oscillatory
  network activity., Frontiers in Cellular Neuroscience 9 (2015) 477.

\bibitem{liu2017effects}
B.~Liu, X.~Chen, C.~Yang, J.~Wu, X.~Gao, Effects of transcranial direct current
  stimulation on steady-state visual evoked potentials, in: Engineering in
  Medicine and Biology Society (EMBC), 2017 39th Annual International
  Conference of the IEEE, IEEE, 2017, pp. 2126--2129.

\bibitem{haberbosch2019rebound}
L.~Haberbosch, S.~Schmidt, A.~Jooss, A.~K{\"o}hn, L.~Kozarzewski,
  M.~R{\"o}nnefarth, M.~Scholz, S.~A. Brandt, Rebound or entrainment? the
  influence of alternating current stimulation on individual alpha, Frontiers
  in human neuroscience 13 (2019) 43.

\bibitem{duan2016effects}
R.~Duan, D.~Zhang, Effects of transcranial alternating current stimulation on
  performance of ssvep-based brain-computer interface, in: 2016 IEEE
  International Conference on Real-time Computing and Robotics (RCAR), IEEE,
  2016, pp. 539--542.

\bibitem{schmidt2013progressive}
S.~Schmidt, A.~Mante, M.~Ronnefarth, R.~Fleischmann, C.~Gall, S.~A. Brandt,
  Progressive enhancement of alpha activity and visual function in patients
  with optic neuropathy: A two-week repeated session alternating current
  stimulation study, Brain Stimulation 6~(1) (2013) 87--93.

\bibitem{berger2018brain}
A.~Berger, N.~H. Pixa, F.~Steinberg, M.~Doppelmayr, Brain oscillatory and
  hemodynamic activity in a bimanual coordination task following transcranial
  alternating current stimulation (tacs): A combined eeg-fnirs study, Frontiers
  in Behavioral Neuroscience 12.

\end{thebibliography}

\section*{Figure captions}

\end{document}